\documentclass{hcig}

\usepackage{amsmath}
\usepackage{graphicx}
\usepackage{hyperref}
\usepackage{algorithm}
\usepackage{algpseudocode}
\usepackage{amsfonts,amssymb}
\usepackage{mathtools}
\usepackage{bm}
\usepackage{wrapfig}
\usepackage{xcolor}
\usepackage[nameinlink]{cleveref}
\usepackage{array,tabularx,booktabs}

\newcolumntype{C}{>{\centering\arraybackslash}X}
\newcolumntype{W}{>{\centering\arraybackslash\hsize=1.2\hsize}X}
\newcolumntype{S}{>{\centering\arraybackslash\hsize=.9\hsize}X}

\newcommand{\R}{\mathbb{R}}

\newcommand{\Dsf}{\mathsf{D}}

\newcommand{\varphibm}{\bm{\varphi}}

\newcommand{\mbm}{\bm{m}}
\newcommand{\zbm}{\bm{z}} 
\newcommand{\ybm}{\bm{y}}

\newcommand{\nbm}{\bm{n}}
\newcommand{\xbm}{\bm{x}}

\newcommand{\Ibm}{\bm{I}}

\algrenewcommand\algorithmicrequire{\textbf{Input:}}
\algrenewcommand\algorithmicensure{\textbf{Output:}}

\title{PRISM: Probabilistic and Robust Inverse Solver with Measurement-Conditioned Diffusion Prior for Blind Inverse Problems}

\author{Yuanyun Hu$^{1,2}$, Evan Bell$^1$, Guijin Wang$^2$, and Yu Sun$^{1,}$\textsuperscript{\Letter}}
\address{$^1$Johns Hopkins University \quad $^2$Tsinghua University\\\smallskip
{\footnotesize \textsuperscript{\Letter}Corresponding author: ysun214@jh.edu}}

\headertitle{PRISM}
\headerauthors{Hu et al.}

\begin{document}

\maketitle
\thispagestyle{firstpagestyle}

\blfootnote{Yuanyun Hu is a student in the Tsinghua–JHU-BME Dual Degree Program. The work was done when Yuanyun Hu studied at Johns Hopkins University. Evan Bell is supported by the U.S. Department of Energy Computational Science Graduate Fellowship.}

\begin{abstract}
Diffusion models are now commonly used to solve inverse problems in computational imaging. 
However, most diffusion-based inverse solvers require complete knowledge of the forward operator to be used. 
In this work, we introduce a novel \textit{probabilistic and robust inverse solver with measurement-conditioned diffusion prior (PRISM)} to effectively address \textit{blind} inverse problems.
PRISM offers a technical advancement over current methods by incorporating a powerful measurement-conditioned diffusion model into a theoretically principled posterior sampling scheme. 
Experiments on blind image deblurring validate the effectiveness of the proposed method, demonstrating the superior performance of PRISM over state-of-the-art baselines in both image and blur kernel recovery.
\end{abstract}

\section{Introduction}

In computational imaging, reconstructing an image when the forward model is partially \textit{unknown} remains a significant challenge.
Common examples of such unknowns in real applications include the sensitivity map in magnetic resonance image (MRI) reconstruction~\cite{Pruessmann.etal1999, Hu.etal2024spicer}, accurate view angles in X-ray computed tomography (CT)~\cite{Basu.etal2000, Xie.etal2021joint}, and the blur kernel in image deblurring~\cite{freemancamerashake, Chen.etal2019blind}.
Mathematically, the reconstruction task can be formulated as a \textit{blind inverse problem}
\begin{equation}
    \ybm = H_{\varphibm}\xbm + \nbm, \quad \nbm \sim \mathcal{N}(\mathbf{0}, \sigma_{\ybm}^2 \Ibm),
\end{equation}
where $\xbm \in \R^n$ is the true underlying signal, $\ybm \in \R^m$ is the observed measurement, $H_{\varphibm}\in\R^{m\times n}$ is the forward matrix with unknown parameters $\varphibm \in \R^p$, and $\nbm$ is Gaussian measurement noise. 
The goal here is to recover $\xbm$ as accurately as possible from $\ybm$, which typically involves jointly estimating $\varphibm$.

Diffusion models have recently emerged as an effective tool for solving inverse problems~\cite{Chung.etal2023diffusion,Kawar.etal2022denoising,Wang.etal2022zero}.
Several methods have been developed for blind inverse problems using diffusion models, such as \textit{BlindDPS}~\cite{chung2023parallel}, \textit{GibbsDDRM}~\cite{murata2023gibbsddrm}, and \textit{Kernel-Diff}~\cite{sanghvi2024kernel}. 
Despite their effectiveness, each of these methods comes with certain limitations.
For example, BlindDPS uses an \textit{unconditional} diffusion model as a prior for estimating $\varphibm$, leaving the rich information about $\varphibm$ contained in $\ybm$ unutilized.
Similarly, although GibbsDDRM is more theoretically principled, it relies on a simple Laplace prior for $\varphibm$, which often results in unsatisfactory recovery.
Finally, while Kernel-Diff uses a conditional diffusion model for estimating $\varphibm$ from $\ybm$, it resorts to using a traditional non-blind deep network to predict $\xbm$ from $\ybm$ for a fixed estimate of $\varphibm$. Since the diffusion model is not used as an image prior, any inaccuracies in estimating $\varphibm$ may propagate into substantial errors in the estimate of $\xbm$.

In this work, we develop a novel diffusion-based blind inverse solver termed \textit{probabilistic and robust inverse solver with measurement-conditioned diffusion prior (PRISM)} by extending the \textit{Plug-and-Play Diffusion Models (PnP-DM)} framework introduced in~\cite{Wu.etal2024} to the blind setting. 
The resulting approach overcomes the major limitations of BlindDPS, GibbsDDRM, and Kernel-Diff while retaining the best features of each of these algorithms. 
In particular, the proposed PRISM uses diffusion models as priors for reconstructing both $\xbm$ and $\varphibm$ (as in BlindDPS), employs a theoretically principled sampling scheme (as in GibbsDDRM), and leverages a powerful \textit{conditional} diffusion model to effectively estimate $\varphibm$ (as in Kernel-Diff).
We note that a similarly derived method, \textit{Blind-PnPDM}~\cite{li2025plug}, was recently proposed; however we demonstrate that PRISM's use of a \textit{measurement-conditioned} kernel prior offers substantial improvements over Blind-PnPDM in terms of both performance and robustness.

\begin{figure*}[t!]
    \centering
    \includegraphics[width=0.95\textwidth]{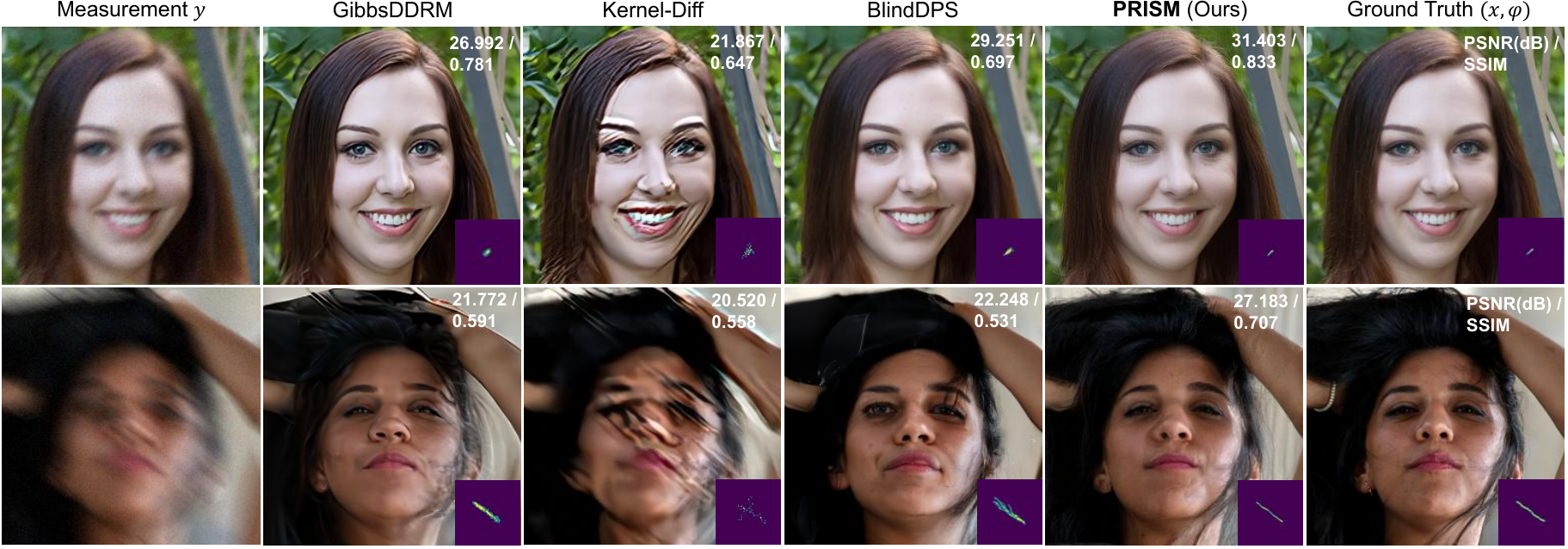}
    \caption{Visual comparison of the image ($\xbm$) and kernel ($\varphibm$) reconstructions obtained by PRISM and baseline methods for the blind motion deblurring task ($\sigma=0.05$). Each reconstruction shown is a single sample generated by each method (single-sample estimation). Note how PRISM recovers fine textures such as hair and skin wrinkles while also retaining high PSNR and SSIM values.
    }
    \label{fig:compare}
\end{figure*}

\section{Method}
\label{sec:method}

\subsection{Blind Bayesian Inverse Problem}
\label{ssec:preliminaries}
From a Bayesian perspective, solving blind inverse problems can be viewed as inferring the joint posterior distribution
\begin{equation}
    p(\xbm, \varphibm \mid \ybm) \propto p(\ybm \mid \xbm, \varphibm) \, p(\xbm) \, p(\varphibm),
\end{equation}
where the likelihood $p(\ybm \mid \xbm, \varphibm)$ enforces data fidelity, $p(\xbm)$ is the prior of the image, and $p(\varphibm)$ is the prior of the unknown parameters. 
In the negative log domain, we denote $g(\xbm,\varphibm; \ybm) = -\log p(\ybm \mid \xbm, \varphibm)$, $r_{\xbm}(\xbm) = -\log p(\xbm)$, and $r_{\varphibm}(\varphibm) = -\log p(\varphibm)$, giving
\begin{equation}
    \label{eq:joint_posterior}
    p(\xbm,\varphibm \mid \ybm) \propto \exp\!\left(-g(\xbm,\varphibm; \ybm) - r_{\xbm}(\xbm) - r_{\varphibm}(\varphibm) \right).
\end{equation}
Directly sampling $p(\xbm, \varphibm \mid \ybm)$ is difficult due to the strong coupling between $\xbm$ and $\varphibm$ and the nonconvex nature of the problem.
To overcome this, we adapt the PnP-DM framework~\cite{Wu.etal2024}, which utilizes a \textit{split Gibbs sampling (SGS)} strategy~\cite{vono2019split} to decouple the likelihood and priors.
This is accomplished by introducing auxiliary variables $\zbm \in \mathbb{R}^n$ and $\mbm \in \mathbb{R}^p$ for $\xbm$ and $\varphibm$ and considering the augmented distribution
\begin{align}
    \label{eq:aug_posterior}
    \pi(\xbm, \zbm, \varphibm, \mbm \mid \ybm)
    &\propto \exp\bigg(
        - g(\zbm,\mbm; \ybm) - r_{\xbm}(\xbm) - r_{\varphibm}(\varphibm) \nonumber\\
    &\quad - \frac{\|\xbm - \zbm\|_2^2}{2\rho_{\xbm}^2}
           - \frac{\|\varphibm - \mbm\|_2^2}{2\rho_{\varphibm}^2}
    \bigg),
\end{align}
where $\rho_{\xbm}, \rho_{\varphibm}>0$ control the coupling between $(\xbm,\zbm)$ and $(\varphibm,\mbm)$. 
The advantage of sampling from this distribution, rather than directly from $p(\xbm, \varphibm \mid \ybm)$, is that the resulting updates for $\xbm$ and $\varphibm$ only involve the prior, while the corresponding updates for $\zbm$ and $\mbm$ only involve the likelihood. 
This strategy is analogous to the \textit{variable splitting} technique used in Half-Quadratic Splitting~\cite{HQS} and ADMM~\cite{admm}, which has been shown to be effective for non-convex problems~\cite{Wang.etal2019}.

\subsection{Proposed Method: PRISM}
\label{ssec:PRISM}

The proposed PRISM aims to sample from~\eqref{eq:aug_posterior} by alternating between four conditional updates, which are outlined in the remainder of this section. 
In each update step, only one variable is updated while the others are kept fixed.

\begin{algorithm}[t]
  \caption{PRISM}
  \label{alg:blind-solver}
  \begin{algorithmic}[1]
    \Require Initialization $\xbm^{0}, \mbm^{0}$; total iterations $K$; coupling schedules $\{\rho_{\xbm}^k\}_{k=1}^{K}, \{\rho_{\varphibm}^k\}_{k=1}^{K}$; likelihood potential $g(\,\cdot\,;\ybm)$; pretrained image diffusion model $\Dsf^{\xbm}(\,\cdot\,)$ (unconditional) and kernel diffusion model $\Dsf^{\varphibm}(\,\cdot\,;\ybm)$ (conditional on $\ybm$)
    \Ensure $(\xbm^{K},\varphibm^{K})$ as an approximate sample from $\pi(\xbm, \varphibm \mid \ybm)$.
    \For{$k = 1,2,\dots,K$}
      \State $\varphibm^{k} \leftarrow \texttt{KernelCondPrior}\!\left(\mbm^{k-1}, \rho_{\varphibm}^{k}, \Dsf^{\varphibm}(\,\cdot\,;\ybm)\right)$
      \State $\zbm^{k} \leftarrow \texttt{ImageLikelihood}\!\left(\xbm^{k-1}, \varphibm^{k}, \rho_{\xbm}^{k}, g(\,\cdot\,;\ybm)\right)$
      \State $\xbm^{k} \leftarrow \texttt{ImagePrior}\!\left(\zbm^{k}, \rho_{\xbm}^{k}, \Dsf^{\xbm}(\,\cdot\,)\right)$
      \State $\mbm^{k} \leftarrow \texttt{KernelLikelihood}\!\left(\xbm^{k}, \varphibm^{k}, \rho_{\varphibm}^{k}, g(\,\cdot\,;\ybm)\right)$
    \EndFor
  \end{algorithmic}
\end{algorithm}

\smallskip
\noindent\textbf{Kernel Conditional Prior Step.} Given $\mbm$, the kernel $\varphibm$ is updated by sampling the distribution
\begin{equation}
\label{eq:phi-step}
p(\varphibm \mid \mbm)
\propto \exp\!\left(
    - r_{\varphibm}(\varphibm)
    - \frac{\|\varphibm - \mbm\|_2^2}{2\rho_{\varphibm}^2}
\right).
\end{equation}
The key insight in the PnP-DM framework is that right hand side of~\eqref{eq:phi-step} is the likelihood of a Gaussian denoising problem with noise variance $\rho_{\varphibm}^2$, prior $p(\varphibm)$, and noisy observation $\mbm$. Explicitly, this can be seen by rewriting the right hand side as
\begin{equation}
    \exp\!\left(
    - r_{\varphibm}(\varphibm)
    - \frac{\|\varphibm - \mbm\|_2^2}{2\rho_{\varphibm}^2}
\right) \propto p(\varphibm)\mathcal{N}(\varphibm; \mbm, \rho_{\varphibm}^2\Ibm).
\end{equation}
It is then straightforward to sample from this distribution using a diffusion model. In particular, \cite{Wu.etal2024} showed that one can simply initialize the reverse process with $\varphibm$ (up to a scaling factor) and then run the reverse process from an appropriately chosen time point corresponding to a noise level of $\rho_{\varphibm}$. We refer to~\cite{Wu.etal2024} for complete details of how this can be performed with arbitrary diffusion models. In PRISM, we implement this step with a \textit{measurement-conditioned} diffusion model denoted by $\Dsf^{\varphibm}(\,\cdot\,; \text{$\ybm$})$.

Empirically, we found that measurement conditioning is not a minor design choice, but is critical to the success of the proposed approach. 
This is demonstrated in Fig.~\ref{fig:convergence}, where we show that Blind-PnPDM, which is conceptually similar to PRISM but uses an unconditional kernel prior, yields poor convergence and fails to find a reasonable solution when $\varphibm$ is initialized randomly.

\smallskip
\noindent\textbf{Image Likelihood Step.} Given $\xbm$ and $\varphibm$, the latent $\zbm$ is drawn from
\begin{equation}
\label{eq:z-step}
p(\zbm \mid \ybm, \varphibm, \xbm) \propto
\exp\!\left(
    -\frac{\|H_{\varphibm} \zbm - \ybm\|_2^2}{2\sigma_{\ybm}^2}
    -\frac{\|\xbm - \zbm\|_2^2}{2\rho_{\xbm}^2}
\right).
\end{equation}
When $H_{\varphibm}$ is linear, this distribution is Gaussian with covariance and mean specified by
\begin{equation}
\label{eq:z-closed-form}
\bm{\Sigma}_{\zbm}^{-1} = \frac{1}{\sigma_{\ybm}^2} H_{\varphibm}^\top H_{\varphibm} + \frac{1}{\rho_{\xbm}^2}I, \quad
\mu_z = \bm{\Sigma}_{\zbm}\!\left( \frac{1}{\sigma_{\ybm}^2} H_{\varphibm}^\top \ybm + \frac{1}{\rho_{\xbm}^2} \xbm \right).
\end{equation}
Importantly, all of the operations in~\eqref{eq:z-closed-form} can be implemented efficiently using the Fast Fourier Transform (FFT). 
For non-linear $H_{\varphibm}$ or non-Gaussian noise, gradient-based MCMC methods such as Langevin dynamics can be used to effectively draw samples~\cite{Laumont.etal2022, Wu.etal2024, Sun.etal2024}.

\smallskip
\noindent\textbf{Image Prior Step.} Given $\zbm$, the image $\xbm$ is updated by sampling from
\begin{equation}
    p(\xbm \mid \zbm) \propto \exp\left( - r_{\xbm}(\xbm) - \frac{\|\xbm - \zbm\|_2^2}{2\rho_{\xbm}^2} \right).
\end{equation}
The implementation of this step is essentially identical to the \textit{Kernel Prior Step}, and is achieved by running the reverse process of a pretrained diffusion model, which we denote by $\Dsf^{\xbm}(\,\cdot\,)$.

\smallskip
\noindent\textbf{Kernel Likelihood Step.} Given $\xbm$ and $\varphibm$, the kernel $\mbm$ is drawn from
\begin{equation}
\label{eq:m-step}
p(\mbm \mid \ybm, \varphibm, \xbm) \propto 
\exp\!\left( -\frac{\|H_{\mbm} \xbm - \ybm\|_2^2}{2\sigma_{\ybm}^2} 
             -\frac{\|\mbm - \varphibm\|_2^2}{2\rho_{\varphibm}^2} \right).
\end{equation}
As in the image likelihood step, this distribution is Gaussian, and its mean and covariance are available in closed form. 
This can be seen by using the fact that convolution is commutative to treat $\xbm$ as the kernel and $\mbm$ as the signal. 
In matrix form, we can write $H_{\mbm}\xbm=C_{\xbm}\mbm$, where $C_{\xbm}$ is an appropriate Toeplitz matrix constructed from $\xbm$. We then obtain the covariance and mean exactly as in the image likelihood step
\begin{equation}
\bm{\Sigma}_{\mbm}^{-1} = \frac{1}{\sigma_{\ybm}^2}C_{\xbm}^\top C_{\xbm} + \frac{1}{\rho_{\varphibm}^2}I, \quad \bm{\mu}_{\mbm} = \bm{\Sigma}_{\mbm}\left( \frac{1}{\sigma_{\ybm}^2}C_{\xbm}^\top\ybm + \frac{1}{\rho_{\varphibm}^2}\varphibm \right).
\end{equation}
More details on how to efficiently compute these quantities and sample from $\mathcal{N}(\bm{\mu}_{\mbm}, \bm{\Sigma}_{\mbm})$ can be found in Appendix C of~\cite{Wu.etal2024}.

\smallskip\noindent
The complete PRISM procedure is summarized in Algorithm~\ref{alg:blind-solver}.  
We initialize $\xbm^{0}$ and $\mbm^{0}$, then iterate the four updates with coupling parameters $\rho_{\xbm}^{k}, \rho_{\varphibm}^{k}$ annealed from large to small values.  
This annealing accelerates chain mixing and helps escape poor local minima in highly ill-posed blind inverse problems.

\section{Numerical Validation}
\label{sec:numerical}

\begin{table}[t]
\centering
\caption{Numerical results obtained by PRISM and baselines for single-sample estimation. All values are averaged over the test dataset. RMSE values are in $10^{-3}$ units. \textbf{Bold} marks best results; \underline{underlined} numbers indicate second best.}
\label{tab:single}
\small
\begin{tabularx}{0.9\columnwidth}{l *{3}{C} @{\hspace{2mm}} *{2}{C}}
\toprule
 & \multicolumn{3}{c}{\textbf{Image}} & \multicolumn{2}{c}{\textbf{Kernel}}\\
\cmidrule(lr){2-4}\cmidrule(l){5-6}
\textbf{$\sigma=0.05$} & PSNR $\uparrow$ & SSIM $\uparrow$ & LPIPS $\downarrow$ & RMSE $\downarrow$ & SSIM $\uparrow$\\
\midrule
GibbsDDRM  & \underline{24.990} & \underline{0.737} & \underline{0.231} & \underline{1.621} & \underline{0.995} \\
BlindDPS   & 24.561 & 0.555 & 0.270 & 2.202 & 0.985 \\
Kernel-Diff & 20.086 & 0.527 & 0.415 & 2.240 & 0.987 \\
\textbf{PRISM} & \textbf{27.317} & \textbf{0.744} & \textbf{0.225} & \textbf{0.788} & \textbf{0.999} \\
\midrule
\textbf{$\sigma=0.02$} & PSNR $\uparrow$ & SSIM $\uparrow$ & LPIPS $\downarrow$ & RMSE $\downarrow$ & SSIM $\uparrow$\\
\midrule
GibbsDDRM  & \underline{25.751} & \underline{0.758} & \underline{0.214} & \underline{1.598} & \underline{0.995} \\
BlindDPS   & 25.597 & 0.598 & 0.246 & 2.202 & 0.985 \\
Kernel-Diff & 20.377 & 0.525 & 0.404 & 2.167 & 0.989 \\
\textbf{PRISM} & \textbf{27.962} & \textbf{0.770} & \textbf{0.209} & \textbf{0.792} & \textbf{0.999} \\
\bottomrule
\end{tabularx}
\end{table}

\begin{table}[t]
\centering
\caption{Numerical results obtained by PRISM and baselines for posterior mean estimation. 
All values are averaged over the test dataset.
RMSE values are in $10^{-3}$ units. \textbf{Bold} marks best results; \underline{underlined} numbers indicate second best.}
\label{tab:mean20}
\small
\begin{tabularx}{0.9\columnwidth}{l *{3}{C} @{\hspace{2mm}} *{2}{C}}
\toprule
 & \multicolumn{3}{c}{\textbf{Image}} & \multicolumn{2}{c}{\textbf{Kernel}}\\
\cmidrule(lr){2-4}\cmidrule(l){5-6}
\textbf{$\sigma=0.05$}& PSNR $\uparrow$ & SSIM $\uparrow$ & LPIPS $\downarrow$ & RMSE $\downarrow$ & SSIM $\uparrow$\\
\midrule
GibbsDDRM  & \underline{28.053} & \underline{0.824} & \underline{0.247} & 2.092 & 0.987 \\
BlindDPS   & 26.706 & 0.779 & 0.295 & \underline{0.838} & \underline{0.998} \\
Kernel-Diff & 23.136 & 0.665 & 0.428 & 1.936 & 0.991 \\
\textbf{PRISM} & \textbf{29.341} & \textbf{0.837} & \textbf{0.194} & \textbf{0.768} & \textbf{0.999} \\
\midrule
\textbf{$\sigma=0.02$}& PSNR $\uparrow$ & SSIM $\uparrow$ & LPIPS $\downarrow$ & RMSE $\downarrow$ & SSIM $\uparrow$\\
\midrule
GibbsDDRM  & \underline{28.238} & \underline{0.836} & \underline{0.233} & 2.087 & 0.987 \\
BlindDPS   & 27.692 & 0.805 & 0.273 & \underline{0.905} & \underline{0.997} \\
Kernel-Diff & 23.527 & 0.676 & 0.411 & 1.874 & 0.991 \\
\textbf{PRISM} & \textbf{29.736} & \textbf{0.845} & \textbf{0.205} & \textbf{0.766} & \textbf{0.999} \\
\bottomrule
\end{tabularx}
\vspace{-10pt}
\end{table}

\subsection{Experimental setup}
\label{ssec:Experimentalsetup}

\begin{figure}[t]
  \centering
  \includegraphics[width=0.65\linewidth]{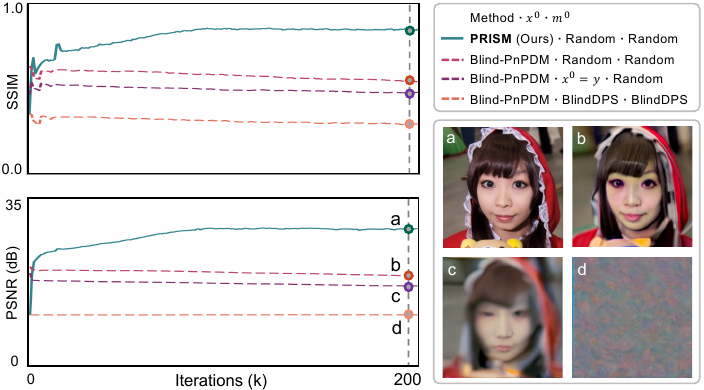}
  \caption{Convergence comparison between PRISM and Blind-PnPDM~\cite{li2025plug}. Results for Blind-PnPDM are shown for three different initialization settings. Note that Blind-PnPDM struggles to converge to a reasonable image with fully random initializations, while PRISM achieves steady convergence to a high-quality image.
}
  \label{fig:convergence}
  \vspace{-10pt}
\end{figure}

We validate PRISM on the task of blind motion deblurring using the FFHQ dataset~\cite{Karras.etal2019}.
% In our experiments, the degradation matrix $H_{\varphibm}$ corresponds to the convolution with blur kernel $\varphi$.
We further corrupt the measurements with the additive white Gaussian noise of standard deviation $\sigma$.
For the image prior, we use the pretrained unconditional image diffusion model from~\cite{Wu.etal2024} as $\Dsf^{\xbm}(\,\cdot\,)$.
For the kernel prior, we implement a measurement-conditioned kernel diffusion model $\Dsf^{\varphibm}(\,\cdot\,;\ybm)$ based on the architecture in~\cite{Saharia.etal2023}.
To train the kernel diffusion model, we use a motion blur kernel generator\footnote{https://github.com/LeviBorodenko/motionblur} to create a dataset comprising $25$ million $(\varphibm, \ybm)$ pairs.
The kernel diffusion model is trained for $500,000$ steps with a batch size of $128$.
For evaluation, we create a test dataset containing $50$ randomly selected FFHQ images and $50$ motion blur kernels.
During inference, the coupling parameters $\rho_{\xbm}^{k}$ and $\rho_{\varphibm}^{k}$ are exponentially annealed.
All hyperparameters of PRISM are fine-tuned using a separate validation dataset. 
Detailed hyperparameter values and model architectures are provided in the code\footnote{Code will be released upon paper acceptance.}.
To measure the image reconstruction quality, we use \textit{peak signal-to-noise ratio (PSNR)} and \textit{structural similarity index measure (SSIM)}. 
We also employ learned \textit{perceptual image patch similarity (LPIPS)} for quantifying the human perception quality.

\subsection{Experimental Results}
\label{ssec:results}
\textbf{Reconstruction Performance.} 
We evaluate the reconstruction performance of PRISM for both the image and the blur kernel.
We compare PRISM with three state-of-the-art baselines: GibbsDDRM~\cite{murata2023gibbsddrm}, BlindDPS~\cite{chung2023parallel}, and Kernel-Diff~\cite{sanghvi2024kernel}.
In particular, we report the numerical results for two scenarios: (i) \textit{single-sample estimate} (Table~\ref{tab:single}), which considers only one posterior sample $(\xbm,\varphibm)$ for each method; and (ii) \textit{posterior mean estimate} (Table~\ref{tab:mean20}), where PRISM averages 20 posterior samples $\{(\xbm_i,\varphibm_i)\}^{20}_{i=1}$ from one converged chain while baselines average the output of 20 independent runs (see further explanation in \textit{Uncertainty Quantification}). 
The first scenario matches more real-world use cases, while the second one aims to approximate the mean of the posterior for optimal performance in terms of \textit{mean squared error (MSE)} and PSNR.
Fig.~\ref{fig:compare} presents a visual comparison of results obtained by PRISM and the baseline methods.
Across all estimation scenarios and noise levels, PRISM consistently achieves superior numerical performance in both image reconstruction and kernel estimation. 
Notably, PRISM outperforms GibbsDDRM (the best baseline) by more than $2$ dB in PSNR for single-sample estimation.
In addition, PRISM accurately restores the blur kernel, attaining the lowest \textit{root MSE (RMSE)} and highest SSIM values.
The visual results in Fig.~\ref{fig:compare} further demonstrate PRISM's outstanding performance. 
Note the accurate recovery of fine image textures such as hair and skin wrinkles, as well as the blur kernel itself.

\smallskip
\noindent
\textbf{Robustness \& Convergence}.
In this section, we show that the inclusion of the measurement $\ybm$ as a condition in the kernel diffusion prior is critical for ensuring the robustness and convergence of PRISM.
Fig.~\ref{fig:convergence} plots the PSNR and SSIM obtained by PRISM and Blind-PnPDM~\cite{li2025plug} across $200$ iterations; the final images are also visualized for comparison.
We implemented Blind-PnPDM following the pseudocode and hyperparameter configurations provided in~\cite{li2025plug}.
While PRISM is initialized only with random $\xbm^0$ and $\mbm^0$, we consider three different initializations for Blind-PnPDM to ensure full exploration of its potential: \textit{(i)} random $\xbm^0$ and $\mbm^0$, \textit{(ii)} $\xbm^0=\ybm$ and random $\mbm^0$, and \textit{(iii)} the outputs of BlindDPS as $\xbm^0$ and $\mbm^0$. 
As shown in the Fig~\ref{fig:convergence}, PRISM converges to a high-quality image from the random initializations, with steady gains until saturation. 
On the other hand, Blind-PnPDM fails to converge to a reasonable image in this case.
By offering better initializations, we observe that the performance of Blind-PnPDM improves; see images (b) and (c) in Fig~\ref{fig:convergence}.
Nevertheless, it still yields to inferior reconstructions to PRISM and shows unwanted sensitivity to different initializations.
Note how image (a) provides better visual quality compared to images (b), (c), and (d).

\begin{figure}[t!]
  \centering
  \includegraphics[width=0.65\linewidth]{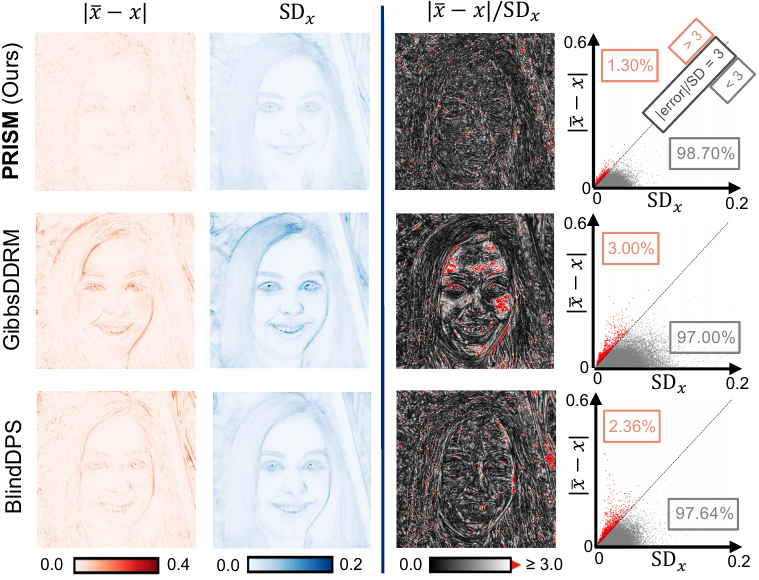}
  \caption{Visualization of the pixel-wise statistics associated with the image reconstruction ($\xbm$) shown in Fig.~\ref{fig:compare} (\textit{1st row}). The left columns plot the absolute error ($|\bar{\xbm}-\xbm|$) and standard deviation (SD), and the right columns plot the 3-SD credible interval with the outlying pixels highlighted in red.}
  \label{fig:uq_x}
  \vspace{-5pt}
\end{figure}

\begin{figure}[t!]
  \centering  
  \includegraphics[width=0.65\linewidth]{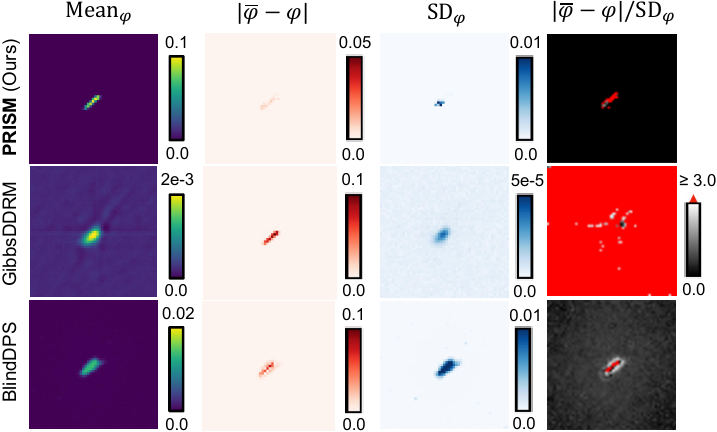}
  \caption{Visualization of the pixel-wise statistics associated with the kernel reconstruction ($\varphibm$) shown in Fig.~\ref{fig:compare} (\textit{1st row}). 
  From left to right, the plots show the sample mean, absolute error ($|\bar{\varphibm}-\varphibm|$), standard deviation (SD), and error-to-SD ratio, where the outlying pixels are highlighted in red.
  }
  \label{fig:uq_k}
  \vspace{-10pt}
\end{figure}

\begin{table}[t]
\centering
\caption{The averaged absolute error ($|\bar{\xbm}-\xbm|$), SD and NLL values obtained by PRISM and baselines for image reconstruction. 
All values are averaged over the test dataset.
\textbf{Bold} marks best results; \underline{underlined} numbers indicate second best.}
\label{tab:uq}
\small
\begin{tabularx}{0.9\columnwidth}{l W S S @{\hspace{3mm}} W S S}
\toprule
 & \multicolumn{3}{c}{\textbf{$\sigma=0.05$}} & \multicolumn{3}{c}{\textbf{$\sigma=0.02$}}\\
\cmidrule(lr){2-4}\cmidrule(l){5-7}
Method & $|\bar{\xbm}-\xbm|$ $\downarrow$ & SD $\downarrow$ & NLL $\downarrow$ & $|\bar{\xbm}-\xbm|$ $\downarrow$ & SD $\downarrow$ & NLL $\downarrow$ \\
\midrule
GibbsDDRM        & \underline{0.026} & \underline{0.029} & \underline{-1.935} & \underline{0.026} & \underline{0.026} & \underline{-1.922} \\
BlindDPS         & 0.030 & 0.030 & -1.861 & 0.027 & 0.027 & \textbf{-2.008} \\
Kernel-Diff       & 0.048 & 0.055 & -0.987 & 0.047 & 0.053 & -1.092 \\
\textbf{PRISM}    & \textbf{0.024} & \textbf{0.023} & \textbf{-1.997} & \textbf{0.023} & \textbf{0.020} & -1.857 \\
\bottomrule
\end{tabularx}
\vspace{-10pt}
\end{table}

\smallskip
\noindent
\textbf{Uncertainty Quantification.} 
We lastly discuss the uncertainty quantification (UQ) enabled by PRISM as a posterior sampling method.
We considered GibbsDDRM~\cite{murata2023gibbsddrm}, BlindDPS~\cite{chung2023parallel}, and Kernel-Diff~\cite{sanghvi2024kernel} as baselines, all of which are based on the reverse diffusion framework.
PRISM differs from these methods by adopting a Markov chain Monte Carlo (MCMC) formulation, allowing samples to be drawn from a single converged chain rather than requiring multiple runs of the algorithm for generating different samples.
To quantitatively measure the quality of UQ, we compute the normalized \textit{negative log-likelihood (NLL)}~\cite{Lakshminarayanan.etal2017} of the ground truth $\xbm$, assuming independent pixel-wise Gaussian distributions characterized by sample mean $\bar{\xbm}$ and standard deviation $\mathsf{SD}$.
Note that better UQ algorithms minimize NLL by producing an accurate $\bar{\xbm}$ and avoiding an excessively large $\mathsf{SD}$.
Table.~\ref{fig:uq_x} summarizes the averaged NLL values obtained by all methods. 
We additionally summarize the pixel-wise absolute error ($|\bar{\xbm}-\xbm|$) and standard deviation ($\mathsf{SD}$) for completeness.
The results show that PRISM achieves competitive UQ performance compared with baselines. 
In particular, PRISM yields more accurate sample mean and avoids large SD.
Fig.~\ref{fig:uq_x} visualizes the pixel-wise statistics associated with the image reconstruction in Fig.~\ref{fig:compare} (\textit{1st row}).
In the right column, we plot the 3-SD credible interval, with outside pixels highlighted in red.
Note that around $99\%$ of the pixels in the ground-truth image lie in the 3-SD interval produced by PRISM, which is superior to that achieved by GibbsDDRM and BlindDPS.
Fig. \ref{fig:uq_k} further visualizes the pixel-wise statistics of the reconstructed motion kernel, including the sample mean, absolute error, SD, and error-SD ratio.
First, GibbsDDRM is overly confident in its inaccurate mean, as evidenced by its large absolute error and excessively small SD. 
While BlindDPS improves the accuracy of the sample mean, it yields large SDs for most pixels in the kernel region.
In contrast, PRISM achieves both an accurate sample mean and a small SD.

\section{Conclusion}
\label{sec:conclusion}

In this work, we introduced PRISM as a novel method for solving blind inverse problems with diffusion models. 
The proposed method is based on split Gibbs sampling, and the resulting algorithm consists of four sampling steps: a \textit{measurement-conditioned} kernel prior step, an image likelihood step, an image prior step, and a kernel likelihood step. The likelihood steps involve closed-form updates that are readily computed, while the image prior step employs a pre-trained image diffusion model and the kernel prior step uses a measurement-conditioned kernel diffusion model.
We empirically validated the effectiveness of PRISM on the blind deblurring task using the FFHQ dataset. 
Experimental results show that PRISM offers improved recovery of both the image and blur kernel over existing state-of-the-art methods. 
Furthermore, PRISM demonstrates strong robustness to initialization. 
Stable convergence to a high-quality image solution is observed even with fully random initialization. 
Additional experiments on UQ further corroborate PRISM's capability as a sampling method to generate reliable samples for both the image and kernel.

% \clearpage

\bibliographystyle{IEEEtran}
\bibliography{references}

\end{document}